\documentclass[seceq]{ptptex}
\usepackage[xdvi]{graphicx}

\notypesetlogo                       

\preprintnumber{%
IT/HEP-552 \\
hep-th/0604121 \\
April, 2006 \\
}
\title{%
Selecting Gauge Theories on an Interval \\
by 5D 
Gauge Transformations
}
\author{%
Norisuke \textsc{Sakai}
\footnote{E-mail: nsakai@th.phys.titech.ac.jp}
and
Nobuhiro \textsc{Uekusa}
\footnote{E-mail: uekusa@th.phys.titech.ac.jp}
}
\inst{%
Department of Physics, Tokyo Institute of Technology, 
Tokyo 152-8551, Japan
}

\recdate{March 9, 2007}
\abst{%
Gauge symmetry breaking by boundary conditions is studied 
in a general warped geometry in five dimensions. 
It has been 
suggested
that a wider class 
of boundary 
conditions 
is
allowed by requiring only vanishing surface 
terms 
when deriving 
the field equations for gauge theories 
on 
an interval (i.e., employing a 
variational principle), 
in comparison
to the twist 
in orbifolding with automorphisms of the Lie algebra. 
We find that there are classes of boundary conditions 
allowed by the variational principle which violate the 
Ward-Takahashi identity and give 
four-point tree 
amplitudes 
that increase 
with energy in channels 
that have not yet been 
explored, leading to cross sections 
that increase
as 
powers of the
energy (
which violates
the tree level 
unitarity). 
We also find that such boundary conditions are forbidden 
by 
the requirement that the definitions of 
the restricted
class of five-dimensional (5D) 
gauge transformations
be consistent. 
}

\begin{document}

\maketitle

\section{Introduction}

In models with extra dimensions, there are various 
possibilities for gauge fields. 
The 
initial proposal for the large extra dimensions assumes 
that all particles in the standard model are localized 
on a brane with 
a
four-dimensional (4D) 
world volume. 
\cite{Arkani-Hamed:1998rs}${}^{-}$\cite{Antoniadis:1990ew}
Formulating the 
localization of gauge fields on a wall is a challenging 
problem, which can be 
realized in certain models. 
\cite{Dvali:1996xe}${}^{,}$ 
\cite{Isozumi:2003uh} 
However, other interesting possibilities arise if the 
gauge fields are propagating in the higher-dimensional 
bulk spacetime. 
The extra-dimensional component of gauge fields can act 
as a Higgs scalar field to break the gauge symmetry 
\cite{Hatanaka:1998yp}. 
The Wilson line dynamics can provide another source of 
gauge symmetry breaking, namely, the Hosotani mechanism 
\cite{Hosotani:1983xw}. 
If the extra dimensions 
are compactified on a topologically 
nontrivial manifold, such as $S^1$, 
twisting can be 
realized, 
and
the Scherk-Schwarz symmetry breaking 
mechanism thereby appears \cite{Scherk:1979zr}. 
The key 
to these mechanisms 
can be summarized as a nontrivial holonomy along 
a nontrivial cycle, 
which can also be understood as 
vacuum expectation values of adjoint scalar fields coming 
from gauge field components along the extra dimensions.
If orbifolds are introduced, one can also impose 
boundary conditions at the fixed points of the orbifold 
to break 
all or part of a gauge group,
usually using the automorphisms 
of the Lie algebra. 
\cite{Kawamura:1999nj}${}^{-}$\cite{Quiros:2003gg}
Combined with the Wilson lines, the orbifold models have 
recently gained much attention. 
\cite{Hall:2001tn}${}^{-}$\cite{Hosotani:2006qp}
A 
class of boundary conditions 
wider
than 
the orbifolding with automorphisms 
has been pursued to obtain more realistic 
models, in particular, to reduce the rank of the gauge 
group \cite{Hebecker:2001jb}. 
One notable proposal 
is 
to consider gauge theories on 
an interval and to require that the surface terms must 
vanish in order for the variational principle 
 \cite{Csaki:2003dt} to give 
field equations.\footnote{
Alternatively, the hermiticity of the supercharges of 
supersymmetric quantum mechanics has also been  proposed 
to determine the boundary conditions \cite{Nagasawa:2004xk}. 
If applied merely at the quadratic level, as in 
Ref.
22),
it also allows the undesirable boundary 
conditions that we found to violate the Ward-Takahashi 
identity and 
tree level unitarity. 
However, imposing 
this condition of hermiticity
at the nonlinear level, 
one finds that 
it
eventually leads 
a
condition 
that is identical to
our condition of the restricted class of 5D 
gauge transformations.
} 
By imposing boundary conditions, part of 
the
five-dimensional 
(5D) gauge 
invariance is explicitly broken, although the 5D 
gauge invariance is 
preserved
in the bulk. 
It is 
commonly believed that
in 4D,
gauge 
invariance is 
vital
to guarantee
the Ward-Takahashi 
identity and 
unitarity. 
Therefore it is 
important to determine
whether or not 
the wider 
range
of boundary conditions allowed by 
the variational principle 
introduces difficulties in regard to these matters.
It should also be useful to examine if 
a
certain restricted 
class of 5D 
gauge transformations (
with gauge transformation functions restricted by 
certain boundary conditions)
can 
be consistently defined as 
the 
invariance of the theories with 
boundaries.

Higgs bosons are needed in 
4D
to cancel 
terms 
that increase
with energy in 
the
scattering amplitudes of 
longitudinal massive gauge bosons \cite{LlewellynSmith:1973ey}. 
The unitarity relation 
between 
$\langle n \vert$ and $\vert n \rangle$ 
corresponding to a single state,
\begin{equation}
-i\langle n \vert (T-T^\dagger)\vert n \rangle =
\sum_m \langle n \vert T^\dagger \vert m \rangle 
\langle m \vert T \vert n \rangle , 
\label{eq:unitarity}
\end{equation}
gives the imaginary part of the elastic 
scattering amplitude as a sum of cross sections of 
various channels, including elastic ($m=n$) 
and
inelastic ones ($m\not=n$). 
Any two-to-two elastic or inelastic scattering amplitudes 
that increase
with energy lead 
to 
an increasing
elastic 
scattering amplitude, 
because the
contributions of each channel in 
Eq.(\ref{eq:unitarity}) 
are
nonnegative. 
In 
this
case, the unitarity bound for the 
elastic scattering amplitude is violated. 
This is called 
tree level unitarity, which can be 
tested even in higher-dimensional gauge 
theories. 
It has been shown that scattering amplitudes 
that increase
with energy are canceled by the 
exchange of 
Kaluza-Klein (KK) gauge bosons in the higher-dimensional 
gauge theories compactified on the torus $S^1$, 
and thus the tree level unitarity 
is maintained, in spite of 
absence of the explicit 
Higgs scalars \cite{SekharChivukula:2001hz}. 
Even with the wider class 
of boundary conditions allowed 
by the variational principle, 
it has been shown that 
the contributions of the 
elastic scattering of longitudinal massive gauge bosons 
in the same
KK excitation level do not 
increase
with energy. 
Therefore, these boundary conditions 
pass
the 
consistency test of the tree level unitarity, at least 
for the contributions from 
elastic scattering 
\cite{Csaki:2003dt}.

The purpose of 
this
paper is to point out that 
the variational principle allows certain 
simple classes of boundary conditions which violate 
the Ward-Takahashi identity 
and
the tree level unitarity.\footnote{
The 
authors 
have
learned that Masaharu Tanabashi also knew of 
examples of boundary conditions which are allowed 
by the variational principle, and 
violate the equivalence theorem, and consequently 
tree level unitarity. 
} 
In contrast to 
previous works
\cite{Abe:2003vg,Csaki:2003dt,SekharChivukula:2001hz,Ohl:2003dp,Muck:2004br},
we compute scattering amplitudes 
in various channels, including the massless modes 
and
the KK modes at various levels, 
in particular
inelastic scattering amplitudes involving different 
excitation levels. 
We 
also show that the 
condition that the definition of 
the restricted class of 5D gauge transformations be consistent
forbids such classes of boundary conditions.

The  5D 
gauge transformations have been examined by 
consistently gauging the global symmetry 
\cite{Hall:2003yc}. 
In 
4D, 
BRST invariance is useful 
as a sophisticated formulation of gauge theories 
after the gauge fixing 
\cite{Becchi:1974md, Kugo:1977zq, Kugo:1981hm}. 
The BRST approach for higher-dimensional gauge theories 
has been used to 
study
the deconstruction 
approach \cite{He:2004zr} 
and orbifold models. 
\cite{Hall:2003yc}
${}^{-}$\cite{Muck:2004br} 
We find that 
the gauge transformations to select boundary conditions can
be consistently defined in terms of
the BRST formulation as well.\footnote{
Because
the problem of defining the 
gauge transformations 
arises
at the tree level rather 
than 
the loop level, 
using the BRST formulation in selecting the 
boundary conditions is 
equivalent to using 
the classical gauge transformations. 
}

In 
\S\ref{sc:bound_cond_var_pr}, 
the variational principle is briefly reviewed. 
In 
\S\ref{sc:scatt_amp}, we compute the scattering 
amplitudes and find 
that
terms 
that violate
the Ward-Takahashi identity 
and the tree level unitarity 
can be reduced to 
functions of the values of the mode 
functions at the boundaries. 
In 
\S\ref{sc:bound_cond_WT}, we demonstrate that the 
variational principle allows 
boundary conditions 
that violate
both the 
Ward-Takahashi identity and 
unitarity. 
We 
refer to
such
boundary conditions as the 
coset-N/subgroup-D boundary conditions. 
In 
\S\ref{sc:brst}, we show that the 
coset-N/subgroup-D boundary conditions do 
not allow 
a consistent definition of 
a restricted class of 5D gauge transformations. 
We also 
discuss
the BRST formulation. 
Useful formulas for mode functions are summarized in 
Appendix \ref{Ap:KK}. 
Some details of the 
computation of scattering amplitudes are given in 
Appendix \ref{Ap:amp}.

\section{Variational principle and scattering amplitudes}
\label{sc:var_pr_sct}

\subsection{Boundary conditions obtained from the variational principle}
\label{sc:bound_cond_var_pr}

To a good approximation, our 4D spacetime is flat 
(except for a 
very small positive cosmological constant). 
Assuming 
flat 4D 
slices in 5D 
spacetime, we parameterize the generic metric in 
terms of two\footnote{
The extra-dimensional coordinate $y$ can be reparameterized 
such that
either one of the two positive 
functions $e^{-4W(y)}$ or $g_{55}(y)$ 
is
constant; 
for example, we could choose
$g_{55}(y)=1$. 
However, we leave these two functions 
unfixed, in order to 
to accommodate various coordinate systems. 
} arbitrary functions, $W(y)$ and $g_{55}(y)$, 
of the extra-dimensional coordinate $x^5 \equiv y$: 
\begin{eqnarray}
 ds^2=g_{MN}dx^Mdx^N=
e^{-4W(y)}\eta_{\mu\nu}dx^\mu dx^\nu+g_{55}(y)dy^2 .  
\end{eqnarray}
Here,
the 
upper-case
Latin indices $M,N,\cdots=0, 1, 2, 3, 5$ 
run over the five spacetime dimensions, $g_{MN}$ is the 
5D 
metric, and the 4D 
spacetime is flat: $\eta_{\mu\nu}=\textrm{diag}(-,+,+,+)$ 
with $\mu, \nu =0, 1,2,3$. 
We assume the extra dimension to 
consistent of
an interval, 
$0 \le y \le \pi R$, and consider appropriate boundary 
conditions at $y=0$ and $y=\pi R$.  

As a simple illustrative example, we 
consider
pure gauge 
theory, and choose the $SU(N)$ gauge group 
in the case that
we need to 
specify the gauge group. 
Introducing the gauge fixing term with the parameter $\xi$, 
we obtain the action in the $R_\xi$ gauge as
\begin{eqnarray}
        \label{eq:lagr}
\mathcal{S}+\mathcal{S}_{\rm GF}=
\int d^4x\,\int_0^{\pi R} dy \sqrt{-g(y)}\left(
-\frac{1}{4} F_{MN}^a F_{PQ}^a g^{MP}g^{NQ}
-\frac{1}{2\xi} (G^a{})^2\right). 
\end{eqnarray}
The field strength is given by 
$F^a_{MN}=\partial_M A_N^a -  \partial_N A_M^a
+g_5 f^{abc}\, A^b_M A^c_N$, where $f^{abc}$ 
is the structure constant of the gauge group, and $g_5$ 
is the 5D gauge coupling. 
We 
choose the gauge fixing function $G^a$ 
so as
to cancel the cross terms 
between the 4D components $A_\mu^a$ and the 
extra-dimensional component $A_5^a$ 
(4D scalar): \cite{Muck:2001yv} 
\begin{eqnarray}
 G^a=e^{4W}\left(\eta^{\mu\nu}\partial_\mu A_\nu^a
   +\xi {1\over \sqrt{g_{55}}}
\partial_5 {e^{-4W}\over \sqrt{g_{55}}}
    A_5^a\right) . 
\end{eqnarray}
We omit 
the corresponding ghost fields,
as
they do not 
appear in tree level scattering amplitudes.

Let us briefly review the variational principle used to 
determine the boundary conditions \cite{Csaki:2003dt}. 
The action~(\ref{eq:lagr}) should 
realize its
minimum with 
respect to variation 
about
the configuration satisfying 
the field equations in the bulk. 
To 
obtain
the field equations, 
we must 
perform 
integrations by parts, 
which results
in the boundary terms 
\begin{eqnarray}
(\delta \mathcal{S}
+\delta \mathcal{S}_{\rm GF})_{\textrm{\scriptsize boundary}}  = 
-\int d^4x \left[ {e^{-4W}\over \sqrt{g_{55}}}
  \eta^{\mu\nu} F^{a}_{5 \mu} \, \delta A_{\nu}^a  
 +{e^{-8W}\over \sqrt{g_{55}}}G^a\delta A_5^a 
 \right]_{y=0}^{y=\pi R} . 
\end{eqnarray}
These boundary terms 
must
vanish 
for the variational principle to be well-defined.
The 
simplest way to satisfy this condition is to use 
the Neumann, $\partial_5 \left. A_{\mu}^a\right|=0$ 
($\partial_5 {e^{-4W}\over \sqrt{g_{55}}} \left. A_5^a\right|=0$), 
or 
Dirichlet, $\left. A_{\mu, 5}^a\right|=0$, boundary 
conditions. 
It has been found that the 
following three 
forms
of the Neumann and/or 
Dirichlet boundary conditions 
provide
the solution: 
 \cite{Csaki:2003dt} 
\begin{eqnarray}
&& \left. A_\mu^a\right|=0, \qquad \left. A_5^a\right|=0, 
\label{eq:var_bnd_cod1}
\\
&& \left. A_\mu^a\right|=0, \qquad 
\partial_5 {e^{-4W}\over \sqrt{g_{55}}} \left. A_5^a\right|=0, 
\label{eq:var_bnd_cod2}
\\
&& \partial_5 \left. A_\mu^a\right|=0, \qquad 
\left. A_5^a\right|=0. 
\label{eq:var_bnd_cod}
\end{eqnarray}
The Neumann conditions $\partial_5 \left. A_\mu^a\right|=0$ 
for the 4D 
vector $A^a_{\mu}$ at both 
boundaries, 
$y=0$ and $\pi R$,
are necessary for the existence of a zero mode 
with 
the color
$a$ 
in the low-energy effective theory. 
It 
is also known
that the choice of these three 
types
of boundary conditions can be made independently 
for each 
color
$a$ and for the 4D 
vector $A^a_{\mu}$ and the 4D 
scalar $A^a_5$. 
The 4D Lorentz invariance also 
implies that the 
gauge generators of the 4D 
massless gauge fields 
$A^a_{\mu}$ 
form a group, as argued in Ref.
21):
\begin{eqnarray}
\partial_5 \left. A_\mu^a\right|=0, \qquad 
a \in H \subseteq G , 
\qquad 
y=0, \pi R ,
\label{eq:gr_bnd_cod}
\end{eqnarray}
where $H$ is a (sub)group 
of the gauge group $G$. 

In the case of the second 
choice of the boundary conditions,
(\ref{eq:var_bnd_cod2}), and the third choice, 
(\ref{eq:var_bnd_cod}), 
mode functions 
for $A_\mu$ and $A_5$ satisfy simple relations, such as 
(\ref{mtb}) in Appendix \ref{Ap:KK}. 
This simplicity
plays an important role 
in 
the investigation of
possible violations of the Ward-Takahashi 
identity and unitarity. 
In the case of the first choice, (\ref{eq:var_bnd_cod1}), 
the mode functions for $A_\mu$ and $A_5$ satisfy a different 
relation, and 
this results in the necessity of
a separate treatment. 
Therefore, we 
consider the second 
and 
third 
choices, (\ref{eq:var_bnd_cod2}) and (\ref{eq:var_bnd_cod}),
in this section and 
discuss the first choice, (\ref{eq:var_bnd_cod1}), only 
briefly in 
\S\ref{sc:brst}. 
We 
denote the Neumann or 
Dirichlet boundary conditions for the 4D 
vector $A_\mu^a$ at each 
boundary with the 
color
$a$ as $D(a)$. 
(
The opposite 
boundary conditions for the scalar $A_5^a$ are implied.) 
Using these boundary conditions, 
we can obtain the 
$n$-th mode functions $f_n^{D(a)}(y)$ 
and $g_n^{D(a)}(y)$ 
for the vector $A_{\mu}^a(x,y)$ 
and scalar  $A_{5}^a(x,y)$, respectively, 
and can decompose 
them into the KK effective 
4D 
fields $A_{\mu n}^a(x)$ 
and $A_{5 n}^a(x)$ as 
\begin{eqnarray}
 A_{\mu}^a(x,y)&=&\sum_{n=0}^{\infty} A_{\mu n}^a(x) f_n^{D(a)}(y), 
\label{eq:kk_vector}
\\ 
A_5^a(x,y)&=&\sum_{n=0}^{\infty} A_{5 n}^a(x) g_n^{D(a)}(y) .
\label{eq:kk_scalar}
\end{eqnarray}
Important properties of the  mode functions are given in 
Appendix~\ref{Ap:KK}. 


\subsection{Scattering amplitudes}
\label{sc:scatt_amp}

\begin{figure}[b]
\begin{center}
\includegraphics{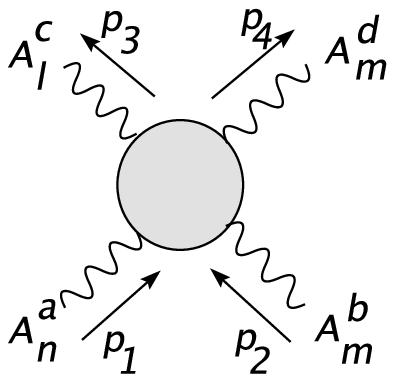}
\caption{$A_n^a A_m^b\to A_l^c A_m^d$ \label{ampfig}}
\end{center}
\end{figure}

Now, we 
examine scattering amplitudes 
for various choices of
Dirichlet 
and/or
Neumann boundary conditions 
at each boundary and for each color.
We consider the gauge boson scattering 
$A_n^a A_m^b\to A_l^c A_m^d$, 
where the subscripts $l, n$ and $m$ 
indicate the
KK levels, 
and the superscripts $a,b,c$ and $d$ indicate the colors, 
as illustrated in Fig.~\ref{ampfig}. 
For 
simplicity, we choose the KK level and boundary 
conditions for the gauge bosons with the 
colors
$b$ and $d$ 
identical. 
In the center-of-mass frame,
the scattering angle and the total energy are denoted as $\theta$ and 
$E\equiv E_n+E_m$, respectively.
To examine the possible growth of scattering amplitudes at a
fixed scattering angle as the total energy $E$ 
increase, 
we choose polarization vectors $\epsilon$ 
for gauge bosons to be longitudinal, except for the 
gauge boson $A_l^c$, as tabulated in Table~\ref{Tab:kin}. 
We 
examine both longitudinal and transverse 
polarizations and massive as well as massless modes 
for $A_l^c$. 
%
%
%
\begin{table}[h]
\caption{Kinematics: The energy of each mode  
is given 
by $E_n=\sqrt{p^2+M_n^2}$. 
Both the longitudinal and transverse polarizations 
$\epsilon(p_3)$ are explored for the gauge 
field $A_l^c$.\label{Tab:kin}}
\begin{tabular}{ll}\hline\hline
 $p_1=(E_n,0,0,p)$  &
        $\epsilon(p_1)=(p/M_n,0,0,E_n/M_n)$ \\
 $p_2=(E_m,0,0,-p)$  &
  $\epsilon(p_2)=(p/M_m,0,0,-E_m/M_m)$ \\ 
 $p_3=(E_l,p'\sin\theta,0,p'\cos\theta)$  &
  $\epsilon(p_3)=$ 
  Eq.(\ref{polart}) (for transverse) or
  Eq.(\ref{polarl}) (for longitudinal)\\
 $p_4=(E_m',-p'\sin\theta,0,-p'\cos\theta)$  &
   $\epsilon(p_4)=(p'/M_m,-E_m'\sin\theta/M_m,0,
   -E_m'\cos\theta/M_m)$ \\
\hline 
 \end{tabular}
\end{table}
Some details of the scattering amplitude calculation 
with arbitrary gauge parameters $\xi$ 
are given 
in Appendix~\ref{Ap:amp}. 
To test the tree level unitarity, we study 
the scattering amplitude at 
energies 
high enough that
the total energy $E$ is much larger 
than any mass 
$M_k$
for the 
intermediate states as well as the external states. 
Leaving the polarization $\epsilon(p_3)$ for $A_l^c$ 
unspecified,  
the terms 
that increase 
with
$E$ in the invariant matrix 
element for the gauge parameter $\xi\ll E^2/M_k^2$ 
are
given by 
\begin{eqnarray}
 && ig_5^2 f^{abe}f^{cde} {E^3\over 8M_n M_m^2}
\epsilon_{\mu}^*(p_3)
 \sum_k \Bigg\{ F_{lmk}^{D(cde)}F_{nmk}^{D(abe)}
 {\cal A}^{\mu}
\nonumber\\
&&
 +{2\over E^2} \left(M_l T_{mkl}^{D(dec)}
-M_m T_{lkm}^{D(ced)}\right)
    \left(M_n T_{mkn}^{D(bea)}-M_m T_{nkm}^{D(aeb)}\right)
      \left(0,-\sin\theta,0,1-\cos\theta\right)^{\mu}
   \Bigg\} ,
\nonumber\\   \label{invm}
\end{eqnarray}
with the four-vector ${\cal A}^{\mu}$ whose components are 
${\cal A}^{\mu=2}= 0$ and 
\begin{eqnarray}
 {\cal A}^{0}
 &\!=\!& 
1+{3M_n^2-3M_m^2+2M_l^2-4M_k^2\over E^2}
    -\left(1+{M_n^2+M_m^2\over E^2}\right)\cos\theta,
\\
 {\cal A}^{1}
 &\!=\!& 
    \left(1+{M_n^2-7M_m^2-2M_l^2+2M_k^2\over E^2}\right)
\sin\theta
\nonumber\\
&&\quad
    -\left(1+{M_n^2+M_m^2-2M_l^2\over E^2}\right)
\sin\theta\cos\theta,
\\
 {\cal A}^{3}
 &\!=\!& 
     {2M_n^2+4M_m^2+2M_l^2-6M_k^2\over E^2}
     +\left(1+{M_n^2-7M_m^2-2M_l^2+2M_k^2\over E^2}\right)
\cos\theta
\nonumber\\
&&\quad
     -\left(1+{M_n^2+M_m^2-2M_l^2\over E^2}\right)
\cos^2\theta . 
\end{eqnarray}
Here,
we have ignored terms which do not 
increase
with energies 
[to ${\cal O}((E/M_k)^0)$]. 
The overlap functions $F_{lmk}^{D(cde)}$ and 
$T_{mkl}^{D(dec)}$ are 
given in Eq.(\ref{overlap}) in Appendix \ref{Ap:KK}. 

Let us consider the case with 
a massive mode for the 
external $A_l^c$ boson. 
Then, the transverse polarization is given by 
\begin{equation}
  \epsilon^*(p_3)=(0,\cos\theta,0,-\sin\theta) 
~~\textrm{or}~~ (0,0,1,0) .
  \label{polart}
\end{equation}
The latter polarization gives a vanishing result upon 
multiplication  
by ${\cal A}^\mu$. 
For the former polarization, the invariant matrix element 
(\ref{invm}) becomes 
\begin{eqnarray}
 && -ig_5^2 f^{abe}f^{cde} {E\sin\theta \over 4M_n M_m^2}{\cal K} ,
   \label{tk}
\end{eqnarray}
where
\begin{eqnarray}
 {\cal K}&=&\sum_k \Bigg\{ \left(M_n^2+2M_m^2-3M_k^2+M_l^2\right)
  F_{lmk}^{D(cde)}F_{nmk}^{D(abe)}
\nonumber\\
&&\qquad
 + \left(M_l T_{mkl}^{D(dec)}-M_m T_{lkm}^{D(ced)}\right)
    \left(M_n T_{mkn}^{D(bea)}-M_m T_{nkm}^{D(aeb)}\right)
   \Bigg\}
\nonumber\\ 
    &=&-3\sum_k \left[{e^{-4W}\over \sqrt{g_{55}}}
       g_k^{D(e)} f_m^{D(b)} f_l^{D(c)}\right]_0^{\pi R}
    \left[{e^{-4W}\over \sqrt{g_{55}}}
       g_k^{D(e)} f_m^{D(d)} f_n^{D(a)}\right]_0^{\pi R}
\nonumber\\
  &&\qquad
    +2\left[{e^{-4W}\over \sqrt{g_{55}}}
       \left(f_n^{D(a)} f_m^{D(b)} f_m^{D(d)} f_l^{D(c)}\right)'
      \right]_0^{\pi R} . 
   \label{kb}
\end{eqnarray} 
For the longitudinal polarization, which is given by
\begin{equation}
 \epsilon^*(p_3)=\left(p'/M_l,E_l \sin\theta/M_l,0,
       E_l \cos\theta/M_l \right) ,   \label{polarl}
\end{equation}
the invariant matrix element is 
\begin{eqnarray}  
 && -ig_5^2 f^{abe}f^{cde} 
{E^2(1-\cos\theta)\over 8M_n M_m^2 M_l}{\cal K} .
    \label{lk} 
\end{eqnarray}
The amplitude 
that increases with energy has the 
coefficient ${\cal K}$ 
for both transverse and longitudinal polarizations. 
We emphasize that the coefficient ${\cal K}$ in 
Eq.~(\ref{kb}) 
reduces to a function of the values of the mode 
functions at the boundaries. 
Therefore, it is directly determined by the choices of 
the boundary conditions. 
If the coefficient ${\cal K}$ of the 
elastic (inelastic)
scattering amplitudes does not vanish, 
the unitarity relation (\ref{eq:unitarity}) gives the 
imaginary part of the elastic amplitude that 
increases 
with energy and violates the unitarity bound.

Equation (\ref{tk}) is applicable 
also for $l\to 0$.
If we 
choose the zero mode ($l=0$) for the gauge boson $A_l^c$, 
the scattering amplitude should satisfy the (on-shell) 
Ward-Takahashi identity: 
The amplitude must vanish if we substitute 
the momentum of the zero mode for 
the polarization vector $\epsilon$. 
Making the substitution 
\begin{equation}
   \epsilon^*(p_3) \to p_3=p'(1,\sin\theta,0,\cos\theta) ,
\end{equation}
we obtain the amplitude (\ref{invm}) as 
\begin{eqnarray}
ig_5^2 f^{abe}f^{cde} (1-\cos\theta)   {E^2\over 8M_n M_m^2}
 f_{l=0}^{D(c)} \left[{e^{-4W}\over \sqrt{g_{55}}}
 (f_m^{D(b)} f_m^{D(d)} f_n^{D(a)}{}'-f_n^{D(a)} f_m^{D(b)}
 f_m^{D(d)}{}')
 \right]_0^{\pi R} ,
\nonumber\\
\label{eq:ward_takahashi_id}
\end{eqnarray}
which 
also reduces 
to a function of the values of the 
mode functions at the boundaries.

We thus find that the violation of both the Ward-Takahashi 
identity and the tree level unitarity are proportional to 
the functions given in (\ref{eq:ward_takahashi_id}) and (\ref{kb}),
which are functions 
of the values of the mode functions at the boundaries.  
In the next subsection, we will 
present explicit
examples 
of boundary conditions that are allowed by the variational 
principle and give nonvanishing functions 
(\ref{eq:ward_takahashi_id}) 
that lead
to the violation 
of the Ward-Takahashi identity and (\ref{kb}) 
that lead to the 
violation of 
tree level unitarity.

\subsection{The coset-N/subgroup-D boundary conditions 
}
\label{sc:bound_cond_WT}

In order to illustrate 
boundary conditions 
that
violate 
the Ward-Takahashi identity and 
tree level unitarity, 
we 
consider the $SU(N)$ gauge group as an example. 
To obtain a semi-realistic symmetry breaking pattern, let us 
consider the symmetry breaking in two steps: 
two different boundary conditions at $y=0$ and $y=\pi R$, 
such as $SU(5) \to SU(3)\times SU(2)\times U(1) \to 
SU(3)\times U(1)$. 
The warp factor $e^{-4W(y)}$ and $g_{55}(y)$ can provide 
two vastly different mass scales for 
the boundaries $y=0$ and $y=\pi R$. 
Without loss of generality, let us take one of the 
boundaries, say $y=0$, to be associated with the high 
energy scale, $M_{\rm GUT}$. 
At this high energy scale, we break the gauge 
group as
\begin{eqnarray}
SU(N) \to SU(N_1)\times SU(N-N_1)\times U(1) 
\label{eq:1st_syM_br}
\end{eqnarray}
by imposing the Neumann boundary conditions 
(\ref{eq:var_bnd_cod}) for the 
4D vectors $A_\mu^a$ 
in the subgroups $SU(N_1)$, $SU(N-N_1)$ and $U(1)$, 
and by imposing the Dirichlet boundary conditions 
(\ref{eq:var_bnd_cod2}) for the coset 
$SU(N)/[SU(N_1)\times SU(N-N_1)\times U(1)]$. 
The boundary conditions 
are written in 
matrix form
as
\begin{eqnarray}
 \left(\begin{array}{l|l}
  \textrm{N} & \textrm{D}\\ \hline
  \textrm{D} & \textrm{N}\\
       \end{array}\right) 
 \!\!\!\!
 \begin{array}{l}
   \} ~\textrm{\scriptsize $N_1$} \\  
   \} ~\textrm{\scriptsize $N-N_1$} \\
 \end{array}
  ~~ \textrm{at}~~ y=0 , \nonumber
\end{eqnarray}
where N and D denote the Neumann and Dirichlet boundary 
conditions, respectively. 
This set of boundary conditions 
violates 
neither
the Ward-Takahashi identity nor 
tree level unitarity. 
It is realizable 
with
automorphisms of the 
Lie algebra in orbifoldings and 
is
often 
used. 
We denote the generators of the subgroups and the coset 
by
\begin{eqnarray}
&&G \in SU(N_1), \qquad 
W \in SU(N-N_1), 
\qquad 
B \in U(1), \nonumber \\
&&X \in SU(N)/[SU(N_1)\times SU(N-N_1)\times U(1)] .
\label{eq:decomp_generators}
\end{eqnarray}

Now we consider the boundary condition at $y=\pi R$, which 
is associated with the lower energy scale, $M_{\rm W}$. 
In order to preserve 
gauge invariance with respect to 
$SU(N_1)$ and $U(1)$, 
we impose the Neumann boundary condition for 
the generators
$G \in SU(N_1)$ and $B \in U(1)$ 
at both 
boundaries, $y=0$ and $\pi R$. 
To break the gauge 
group 
$SU(N-N_1)$ by means of 
boundary conditions, we impose the Dirichlet boundary 
condition (\ref{eq:var_bnd_cod2}) for $W \in SU(N-N_1)$ 
generators. 

The remaining generators belong to the coset 
$X \in SU(N)/[SU(N_1)\times SU(N-N_1)\times U(1)]$. 
Usually, the Neumann boundary conditions 
are not imposed 
on 4D vectors in the coset at both boundaries,
because massless 
gauge fields have to form a group,
as indicated in 
Eq.(\ref{eq:gr_bnd_cod}). 
We have 
the freedom 
of choosing the Neumann boundary conditions here, 
as it is clear that 
the 4D vectors with colors in the coset 
are massive at the high 
energy scale $M_{\textrm{\scriptsize GUT}}$ by the boundary condition 
at $y=0$. 
Thus, we can assign the  boundary condition in 
matrix 
form as 
\begin{eqnarray}
 \left(\begin{array}{l|l}
  \textrm{N} & \textrm{N}\\ \hline
  \textrm{N} & \textrm{D}\\
       \end{array}\right)
 \!\!\!\!
 \begin{array}{l}
   \} ~\textrm{\scriptsize $N_1$} \\  
   \} ~\textrm{\scriptsize $N-N_1$} \\
 \end{array} 
  ~~ \textrm{at}~~ y=\pi R , \nonumber
\end{eqnarray}
with unbroken $U(1)$ 
satisfying the Neumann boundary conditions.
Let us 
refer to
this assignment of the Neumann 
boundary conditions for coset and the Dirichlet boundary 
conditions for a subgroup as the coset-N/subgroup-D 
boundary conditions. 
We 
show that 
these
coset-N/subgroup-D boundary conditions 
violate both the Ward-Takahashi 
identity and 
tree level 
unitarity. 
We 
also show in 
\S\ref{sc:brst}, that this choice 
of boundary conditions 
is forbidden if 
we require the compatibility with 
the restricted class of 5D 
gauge 
transformations.

Nonvanishing values of Eqs.~(\ref{tk}) and (\ref{lk}) 
violate 
tree level unitarity, whereas 
nonvanishing values of 
Eq.~(\ref{eq:ward_takahashi_id}) violate 
the Ward-Takahashi identity. 
Note that 
all of them 
have a common factor of $f^{abe}f^{cde}$. 
Also note that our decomposition in 
Eq.~(\ref{eq:decomp_generators}) 
has the following commutation relations,
as follows from the group 
structure 
\begin{eqnarray}
[G, G] = G, \qquad 
[W, W] = W, 
\qquad 
[X, X] = G + W + B . 
\label{eq:com_rel_generators}
\end{eqnarray}
This implies 
the following nonvanishing structure constants: 
\begin{eqnarray}
f^{GGG}, \quad f^{WWW}, \quad f^{XXG}, \quad 
f^{XXW}, \quad f^{XXB}. 
\label{eq:nonzero_structure_const}
\end{eqnarray}
Therefore, the nonvanishing group theory factor 
$f^{abe}f^{cde}$ can be classified into 
the following three types: among the labels $a,b,c,d$ and $e$ 
the coset generators $X$ appear 
nowhere, 
only in three labels, $(a,c, e), (a,d,e), (b,c,e)$ or $(b,d,e)$, 
or only in four labels, $a,b,c$ and $d$. 

We first examine the violation of the Ward-Takahashi 
identity, which is given by Eq.~(\ref{eq:ward_takahashi_id}) 
and is proportional to 
\begin{eqnarray}
f^{abe}f^{cde} f_{l=0}^{D(c)} \left[{e^{-4W}\over \sqrt{g_{55}}}
 (f_m^{D(b)} f_m^{D(d)} f_n^{D(a)}{}'-f_n^{D(a)} f_m^{D(b)}
 f_m^{D(d)}{}')
 \right]_0^{\pi R} . 
\label{eq:viol_wt_id}
\end{eqnarray}
Let us consider the processes $WX \to BX$ and $WX \to GX$, 
with the zero mode $l=0$ for the massless gauge bosons $B$ 
and $G$, by choosing the colors 
$(a, b, c, d)=(W, X, B, X)$ and $(W, X, G, X)$. 
We find that the 
coset generator $X$ can contribute to the intermediate 
state $e=X$ and that 
the coset-N/subgroup-D boundary conditions 
at $y=\pi R$ give 
\begin{eqnarray}
f^{WXX} f^{BXX}\not=0, \; 
 f_n^{D(a=W)'}\not=0, \; f_m^{D(b=X)}\not=0, \; 
f_{l=0}^{D(c=B, G)}\not=0, \; f_m^{D(d=X)}\not=0. 
\nonumber\\
\label{eq:bc_viol_wt_id}
\end{eqnarray}
The first term 
in Eq.~(\ref{eq:viol_wt_id}) does not vanish, 
whereas the second term vanishes for 
these boundary conditions. 
Therefore, the Ward-Takahashi identity is violated by the 
coset-N/subgroup-D boundary conditions 
in 
these processes, $WX\to BX$ and $WX\to GX$.

We next consider the coefficient 
${\cal K}$ in Eq.~(\ref{kb}), 
which appears in 
Eqs.~(\ref{tk}) and 
(\ref{lk}), for the violation of 
tree level unitarity. 
The coefficient $f^{abe}f^{cde}{\cal K}$ consists of two pieces, 
\begin{eqnarray}
f^{abe}f^{cde} \sum_k \left[{e^{-4W}\over \sqrt{g_{55}}}
       g_k^{D(e)} f_m^{D(b)} f_l^{D(c)}\right]_0^{\pi R}
    \left[{e^{-4W}\over \sqrt{g_{55}}}
       g_k^{D(e)} f_m^{D(d)} f_n^{D(a)}\right]_0^{\pi R} , 
\label{eq:bundary_contr1}
\end{eqnarray}
and 
\begin{eqnarray}
f^{abe}f^{cde} \left[{e^{-4W}\over \sqrt{g_{55}}}
       \left(f_n^{D(a)} f_m^{D(b)} f_m^{D(d)} f_l^{D(c)}\right)'
      \right]_0^{\pi R} .
\label{eq:boundary_contr2}
\end{eqnarray}
Let us choose the same processes, $WX\to BX$ and $WX\to GX$, 
as in the case of the 
violation of the Ward-Takahashi identity, 
without 
restriction to the zero 
mode for $B$ and $G$ in this case. 
For 
these processes, the first term, (\ref{eq:bundary_contr1}), 
vanishes, but the second term, (\ref{eq:boundary_contr2}), 
does not 
vanish,\footnote{
The first term, (\ref{eq:bundary_contr1}), 
is 
nonvanishing with the coset-N/subgroup-D 
boundary conditions if we consider 
the process $XX\to XX$. 
In order to ascertain 
whether
tree level unitarity is 
indeed violated in this process, we need to take account of 
another $s$-channel Feynman diagram, in addition to those 
computed in 
Appendix.\ref{Ap:amp}. 
We 
compute 
inelastic scattering $XX\to XX$ with different KK levels 
for all $X$, 
in contrast to Ref.21),
where only 
the elastic scattering 
is computed with identical 
KK levels for all gauge bosons $X$. 
} because of Eq.~(\ref{eq:bc_viol_wt_id}). 
This establishes that the coset-N/subgroup-D boundary 
conditions violate 
tree level unitarity 
in the processes $WX\to BX$ and $WX\to GX$.

\begin{figure}[t]
\begin{center}
\includegraphics[height=2.5cm]{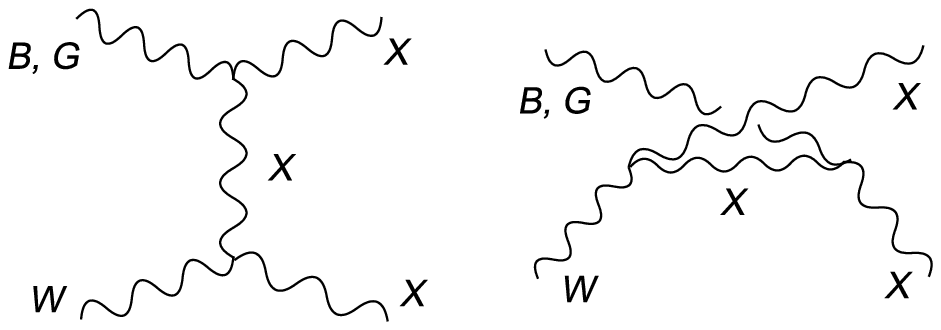}
\end{center}
\caption{Typical Feynman diagrams contributing to the 
violation (\ref{eq:ward_takahashi_id}) of the 
Ward-Takahashi identity and the violation (\ref{kb}) of 
tree level unitarity.
\label{typfd}} 
\end{figure}

We 
illustrate typical Feynman diagrams 
in Fig.\ref{typfd} 
that 
contribute 
to the 
violation (\ref{eq:ward_takahashi_id}) of the 
Ward-Takahashi identity and the violation (\ref{kb}) of 
tree level unitarity. 
The colors of the external gauge bosons with momentum 
assignments in Fig.~\ref{ampfig} are chosen as follows. 
The gauge boson with 
momentum $p_1$ is 
assigned to 
$W \in SU(N-N_1)$, those with $p_2, p_4$ 
to 
$X \in SU(N)/[SU(N_1)\times SU(N-N_1)\times U(1)]$, 
and that with $p_3$ to 
$B \in U(1)$ or $G\in SU(N_1)$. 
We can choose the color of the internal gauge bosons 
to be $X \in SU(N)/[SU(N_1)\times SU(N-N_1)\times U(1)]$.

Thus, we conclude that 
the coset-N/subgroup-D boundary conditions give 
a four gauge boson scattering 
amplitude 
that increases
as powers of the center-of-mass 
energies. 
More specifically, this breakdown of 
tree level unitarity 
occurs if we take one of the external gauge bosons 
in a subgroup with the Dirichlet boundary conditions, 
the other two gauge bosons in the coset 
with the Neumann boundary conditions, and another gauge boson 
in 
an
unbroken subgroup 
with 
the Neumann boundary conditions. 
The Ward-Takahashi identity is also 
broken with this choice of the boundary conditions.

Note that we should not use the unitary gauge 
in the calculation of the tree diagram. 
As we have used the condition $\xi \ll E^2/M_k^2$ in 
computing the scattering amplitude, our calculation is not 
valid for the unitary gauge, 
for which $\xi \rightarrow \infty$. 
The part of the scattering amplitude 
that increase with the energy 
does not depend on the gauge parameter $\xi$ for 
finite values of $\xi$. 
For the unitary gauge, i.e. $\xi= \infty$, the 
extra-dimensional component of the gauge boson 
drops out, as seen from the propagator (\ref{a5prop}). 
Then the breakdown 
of 
tree level unitarity and the 
Ward-Takahashi identity seems to be invisible, even for the 
choice of the coset-N/subgroup-D boundary 
conditions.\footnote{
In 
\S\ref{sc:brst}, we show that the 5D 
gauge 
transformations are ill-defined if the coset-N/subgroup-D 
boundary conditions 
are used. 
Therefore, the different ``gauge choice'' may not 
give the same physical results. 
}

The breakdown 
of 
tree level unitarity and the 
Ward-Takahashi identity 
is a serious problem for 
gauge theories in higher dimensions. 
For this reason,
we look for an additional requirement 
which forbids such a choice of boundary conditions that 
violate 
tree level unitarity and the Ward-Takahashi 
identity in the scattering amplitudes. 
It is most likely that this problem is related to the 
gauge transformation property of gauge fields 
near the boundary. 
We 
examine the 5D 
gauge transformations in the 
next section.

\section{
5D 
gauge transformations and boundary conditions 
}
\label{sc:brst}

\subsubsection*{\underline{\it 5D 
gauge transformations}}

Let us consider 5D 
gauge transformations 
of the 
4D vector component $A_\mu^a(x,y)$,
which are expressed as 
\begin{eqnarray}
\delta A_\mu^a(x,y)=
  \partial_\mu  \epsilon^a(x,y) 
+ g_5 f^{abc} A_\mu^b(x,y) \epsilon^c(x,y) .
\label{eq:gauge_tr_vector}
\end{eqnarray}
The first term 
on the right-hand side is inhomogeneous 
in the gauge field and is independent of the gauge 
coupling constant $g_5$. 
The second nonlinear term is 
first order in $g_5$.
In the bulk, the ordinary 5D gauge invariance is preserved, 
and the transformation function $\epsilon^a$ is an 
arbitrary function of $x$ and $y$.
By contrast,
on the boundaries, $\epsilon^a$, as well as $A_\mu^a$ and $A_5^a$,
is subject to 
certain boundary conditions.
Only if appropriate boundary conditions are chosen 
is the
theory 
invariant under the transformation (\ref{eq:gauge_tr_vector}),
with the transformation function restricted by the boundary conditions.
We call 
such transformations 
the restricted class of 5D gauge transformations.

We first examine the relation between the boundary 
conditions for $A_\mu^a$ and $A_5^a$ to 
leading 
order (i.e., in the limit of a
vanishing gauge coupling, 
$g_5\to 0$). 
The first (inhomogeneous) term 
on the right-hand side of 
Eq.~(\ref{eq:gauge_tr_vector}) 
implies that the 
gauge transformation function $\epsilon^a(x, y)$ 
has the same boundary condition as the corresponding 
4D vector component, $A_\mu^a(x, y)$. 
The gauge transformation of 
the 4D scalar component 
$A_5^a(x,y)$ is given by 
\begin{eqnarray}
\delta A_5^a(x,y)=
  \partial_5  \epsilon^a(x,y) 
+ g_5 f^{abc} A_5^b(x,y) \epsilon^c(x,y) . 
\label{eq:gauge_tr_scalar}
\end{eqnarray}
The first (inhomogeneous) term implies that 
the boundary condition for 
$A_5^a(x,y)$ 
is the same as $\partial_5\epsilon^a$.
In other words, the Neumann (Dirichlet) boundary conditions for 
$A_5^a(x,y)$ 
imply 
the Dirichlet (Neumann) boundary 
conditions for $\epsilon^a(x,y)$. 

\vspace{0.2cm}

\subsubsection*{\underline{\it First choice of the boundary conditions, (\ref{eq:var_bnd_cod1})}}

We can now examine the first choice of the boundary 
conditions, (\ref{eq:var_bnd_cod1}). 
The restricted class of 5D 
gauge transformations 
for $A_\mu^a$ in 
Eq.~(\ref{eq:gauge_tr_vector}) implies that the boundary 
conditions for $A_\mu^a$ and for the gauge transformation 
function $\epsilon^a$ 
are the same, 
whereas the restricted class of 5D 
gauge transformations for $A_5^a$ in 
Eq.(\ref{eq:gauge_tr_scalar}) implies that 
the boundary conditions for $A_5^a$ 
are opposite to 
those for $\epsilon$. 
Therefore, the restricted class of 5D 
gauge transformations 
(at the order of the inhomogeneous term) implies that the 
boundary conditions for $A_\mu^a$ and 
$A_5^a$ 
are opposite.  
The first choice of boundary conditions, 
(\ref{eq:var_bnd_cod1}), allowed by the variational principle 
contradicts\footnote{
The same conclusion 
was reached previously 
with a 
different argument using the unitary gauge \cite{Csaki:2005vy}. 
} 
the gauge 
transformation properties 
of $A_\mu^a(x,y)$ and 
$A_5^a(x,y)$. 
Specifically, the first choice of boundary conditions does not 
allow a consistent definition of the 5D 
gauge transformation 
functions $\epsilon^a(x,y)$.

\vspace{0.2cm}

\subsubsection*{\underline{\it Coset-N/subgroup-D boundary conditions}}

Next, let us study 
whether the coset-N/subgroup-D boundary conditions 
considered in 
\S\ref{sc:bound_cond_WT} 
are consistent with the restricted class of 5D 
gauge transformations 
(\ref{eq:gauge_tr_vector}) and 
(\ref{eq:gauge_tr_scalar}). 
We need to examine the nonlinear terms with the structure 
constant $f^{abc}$ in these gauge transformations. 
The group structure given  in 
Eq.(\ref{eq:nonzero_structure_const}) implies the 
following. 
If $a$ is in a subgroup, 
either both $b$ and $c$ belong to the subgroup 
or both belong to the 
coset. 
If $a$ is in a coset, 
then either $b$ or $c$
belongs to the subgroup and the other 
to the coset. 

Let us consider the restricted class of 5D 
gauge transformations for 
the generators in the coset 
$X \in SU(N)/[SU(N_1)\times SU(N-N_1)\times U(1)]$ with 
the Neumann boundary conditions for $A_\mu^a(x,y)$ (and 
the Dirichlet boundary conditions for $A_5^a(x,y)$). 
The nonlinear term $g_5 f^{abc} A_\mu^b \epsilon^c$ in 
the restricted class of 5D 
gauge transformations  $\delta A_\mu^a$ in 
Eq.~(\ref{eq:gauge_tr_vector}) for 
the 4D vector contains contributions with $A_\mu^b$ 
in the subgroup (coset) and $\epsilon^c$ 
in the coset (subgroup). 
The boundary conditions for $A_\mu^b$ and $\epsilon^b$ with 
the same 
color
$b$ are the same, whereas the 
generators in 
the subgroup and in the coset have 
opposite boundary 
conditions. 
Hence, 
one
of the following two cases 
is realized: 
\begin{eqnarray}
A_\mu^b(x,y=\pi R)=0, 
\quad  
\partial_5 \epsilon^c(x,y=\pi R)= 
0 ,  
\label{eq:cst_N_sub_D1}
\end{eqnarray}
\begin{eqnarray}
\partial_5 A_\mu^b(x,y=\pi R)&=&0, 
\quad  
 \epsilon^c(x,y=\pi R)= 
0 .  
\label{eq:cst_N_sub_D}
\end{eqnarray}
Then the nonlinear term $g_5 f^{abc} A_\mu^b \epsilon^c$ 
consists of a product of functions with the Dirichlet and 
the Neumann boundary conditions: 
\begin{eqnarray}
&&\!\!\!\!\!\!\!\!\!\!\!\!\!\!\!\!\!
\partial_5 \left. \left( A_\mu^b(x,y) 
\epsilon^c(x,y)\right)\right|_{y=\pi R} \nonumber \\
&=&
\partial_5 A_\mu^b(x,y=\pi R)
\epsilon^c(x,y=\pi R)
+A_\mu^b(x,y=\pi R)
\partial_5 \epsilon^c(x,y=\pi R) 
\not=0 . 
\label{eq:prod_D_N}
\end{eqnarray}
Since this nonlinear term does not satisfy the Neumann 
boundary conditions, the boundary conditions for 
$A_\mu^a$ 
are not 
satisfied.

Similarly, we can also examine 
the restricted class of 5D 
gauge transformations for 
the generators in the 
subgroup $W \in SU(N-N_1)$ with 
the Neumann boundary conditions for $A_5^a(x,y)$ 
(and the Dirichlet boundary conditions for 
$A_\mu^a(x,y)$). 
The nonlinear term $g_5 f^{abc} A_5^b \epsilon^c$ in 
the restricted class of 5D 
gauge transformations  $\delta A_5^a$ in 
Eq.~(\ref{eq:gauge_tr_scalar}) for 
4D scalar contains contributions from $b$ and $c$ in the 
coset. 
The boundary conditions in the coset 
are Neumann for the gauge function $\epsilon$ 
and Dirichlet for the 4D scalar $A_5$: 
$A_5^b(x,y=\pi R)=0$, 
$\partial_5 \epsilon^c(x,y=\pi R)=0$. 
Therefore, the product does not 
satisfy the Neumann boundary 
condition. 

In both cases, there is no way to define the gauge 
transformation function $\epsilon^a(x,y)$ consistently. 
Therefore, the coset-N/subgroup-D boundary conditions do not 
allow a consistent definition of 
the restricted class of 5D 
gauge 
transformations.\footnote{
More generally, Dirichlet boundary conditions for $A_\mu^a$ 
in the (non-Abelian) subgroup give 
a similar 
inconsistency for $\delta A_5^a$, although it does not show 
up in Eqs.~(\ref{kb}) and (\ref{eq:ward_takahashi_id}) 
for the 2-to-2 
elastic (inelastic)
scattering amplitudes. 
} 

\vspace{0.2cm}

\subsubsection*{\underline{\it Mode expansions and 4D 
gauge symmetry 
}}

To identify the 4D 
gauge symmetry, 
it is useful to perform mode 
expansions for the 5D 
gauge transformation functions 
$\epsilon^a(x, y)$ as well as the 5D 
gauge fields 
$A^a_M(x, y)$. 
Once the boundary conditions for the gauge transformation 
functions and the gauge fields 
are defined consistently, 
we can perform mode expansions of 
the restricted class of 5D 
gauge transformations of the 
4D vector component $A^a_\mu(x, y)$ in Eq.~(\ref{eq:gauge_tr_vector}) 
as 
\begin{eqnarray}
\sum_{n=0}^{\infty} \delta A_{\mu n}^a(x) f_n^{D(a)}(y)
&=&
\sum_{n=0}^{\infty} \partial_\mu \epsilon_n^a(x) f_n^{D(a)}(y) 
\nonumber \\ 
&+& g_5 f^{abc} \sum_{n=0}^{\infty} A_{\mu n}^b(x) f_n^{D(b)}(y) 
 \sum_{m=0}^{\infty} \epsilon_m^c(x) f_m^{D(c)}(y) .
\label{eq:mode_ex_gauge_tr_v}
\end{eqnarray}
If the zero mode is present for the gauge transformation 
function $\epsilon^a(x, y)$, 
the 
4D gauge invariance is maintained, and the corresponding 
4D gauge field $A^a_{\mu n}(x)$ contains a massless mode. 
If the zero mode is absent for 
 the gauge transformation function $\epsilon^a(x, y)$, 
the 
4D gauge invariance is broken, and
the modes of the corresponding 
4D gauge field 
$A^a_{\mu n}(x)$ are all massive 
with no
zero mode.  

\vspace{0.2cm}

\subsubsection*{\underline{\it Orbifoldings}}

We 
now confirm that the boundary conditions 
realized as the (inner) automorphism of the Lie 
algebra in orbifoldings 
are consistent with 
the restricted class of 5D 
gauge transformations. 
We 
refer to the automorphism at the boundary $y=0$ 
as $P_0$ 
\begin{eqnarray}
 A_\mu (x,-y)=P_0 A_\mu(x,y)P_0^{-1},  
\label{eq:automorphism}
\end{eqnarray}
with $P_0=P_0^\dagger=P_0^{-1}$. 
Similarly, the automorphism $P_1$ is defined at $y=\pi R$. 
To obtain the Dirichlet or Neumann boundary 
conditions, we restrict ourselves to diagonal matrices 
for the automorphisms. 
Then we never obtain 
boundary conditions 
that correspond to the reduction of the 
rank of the group.
\footnote{
More precisely, 
here we are considering 
inner 
automorphisms of the Lie algebra excluding the possibility 
of outer automorphisms \cite{Hebecker:2001jb}. 
}
Combining the automorphism (\ref{eq:automorphism}) 
with the gauge transformation 
\begin{eqnarray}
 \delta A_\mu(x,-y)=\partial_\mu \epsilon (x,-y)
         -ig_5 [A_\mu(x,-y),\epsilon(x,-y)] ,
\end{eqnarray}
we find that the gauge transformation function 
transforms 
under the automorphism 
\begin{eqnarray}
   \epsilon(x,-y)=P_0 \epsilon(x,y) P_0^{-1}. 
\end{eqnarray}
On the other hand, the gauge transformation 
dictates the transformation of $A_5$ under the 
 automorphism 
\begin{eqnarray}
   \delta A_5(x,-y)&=&-\partial_5 \epsilon(x,-y)
                   -ig_5[A_5(x,-y),\epsilon(x,-y)] ,
\\ 
  A_5(x,-y)&=&-P_0 A_5(x,y)P_0^{-1} .
\end{eqnarray}
We can repeat the same analysis at $y=\pi R$. 
This shows that 
a consistent transformation property under the automorphism 
can be assigned for the gauge 
transformation functions. 
This consideration is in accord with the previous result: 
The automorphism of the 
Lie algebra in orbifolding satisfies 
tree level unitarity 
\cite{Ohl:2003dp}.

\vspace{0.2cm}

\subsubsection*{\underline{\it BRST formulation}}

BRST transformations replace the classical gauge 
transformations after the gauge fixing. 
The BRST invariance is 
equivalent to 
the classical gauge invariance and gives no extra 
advantage at the tree level, 
which we are considering. 
We 
briefly consider the BRST 
formulation in order to see that 
we can obtain
the same result 
as in the case of the gauge transformations. 
As long as the boundary conditions for the 
5D
gauge 
transformation function $\epsilon^a$ in 
\begin{eqnarray}
\delta A_M^a(x,y)=
  \partial_M  \epsilon^a(x,y) 
+ g_5 f^{abc} A_M^b(x,y) \epsilon^c(x,y) 
\label{eq:gauge_tr_5D}
\end{eqnarray}
are defined consistently with the boundary conditions for 
the gauge fields $A_M^a$, 
(the restricted class of 5D) 
BRST transformations can be defined 
straightforwardly by promoting the gauge 
transformation function $\epsilon^a(x,y)$ to the ghost 
field $c^a(x,y)$ multiplied by an anti-commuting parameter 
$\varepsilon$. 
We must also add the anti-ghost field $\bar c^a(x,y)$ 
and the Nakanishi-Lautrup field $B^a(x,y)$. 
The restricted class of 5D 
BRST transformations $\delta$ 
are 
given by 
\begin{eqnarray}
 \varepsilon \delta A_M^a(x,y)&=&
  \varepsilon \left[ \partial_M c^a(x,y) 
+g_5 f^{abc}A_M^b(x,y) c^c(x,y)\right] , 
\\
 \varepsilon \delta c^a(x,y)
&=&-{1\over 2} g_5 \varepsilon f^{abc}c^b(x,y) c^c(x,y)  ,
\\
 \varepsilon \delta\bar{c}^a(x,y)&=& i \varepsilon B^a(x,y) ,
\\
 \varepsilon \delta B^a(x,y)&=&0 , 
\end{eqnarray}
where the functions
$c^a(x, y)$ 
satisfy the same boundary conditions as the
corresponding $\epsilon^a(x, y)$. %
\cite{Muck:2001yv}${}^{-}$\cite{Muck:2004br} 

\vspace{0.2cm}

\subsubsection*{\underline{\it Summary and prospects}}

Summarizing our results, we have examined the choice of 
the Neumann 
and Dirichlet boundary conditions to break the 
gauge symmetry in 5D 
gauge theory. 
The variational principle allows 
(coset-N/subgroup-D) 
boundary conditions that violate the Ward-Takahashi 
identity and 
tree level unitarity. 
We have observed that the 
5D 
gauge transformation functions 
$\epsilon(x, y)$, 
in addition to
the gauge fields 
$A_M(x, y)$, 
must
be given 
boundary conditions 
that are 
consistent with the restricted class of 5D 
gauge transformations. 
This condition provides a stringent constraint 
and forbids the coset-N/subgroup-D boundary conditions. 

In considering 
orbifoldings, it is possible to use nondiagonal 
automorphisms of the Lie algebra which give 
boundary conditions that are more involved than Neumann or Dirichlet 
conditions. 
More generally, there are possibilities of nontrivial 
Wilson lines which break the gauge 
group
\cite{Hosotani:1983xw}. 
It has been found that different boundary conditions are 
sometimes related by gauge transformations. 
Under such circumstances, the effective potential at the 1-loop level 
determines  the value of the Wilson lines and 
the symmetry breaking pattern \cite{Haba:2003ux}. 
The elucidation of 
such a possibility is an interesting 
problem. 

We have considered pure gauge theory without matter 
fields in 5D. 
We need to re-analyze the boundary conditions 
in the case that there 
exist 
matter fields, especially if they are localized on the 
boundary. 

The deconstruction approach 
employs
4D gauge theories 
to build higher-dimen-sional gauge models in a discretized 
version \cite{Arkani-Hamed:2001ca}. 
Various boundary conditions, such as the automorphisms 
of the Lie algebra in the orbifolding, 
can be obtained as appropriate limits from the 
deconstruction \cite{He:2004zr}.
It has been shown that 
these boundary conditions 
satisfy 
tree level 
unitarity automatically. 
The deconstruction can provide realistic models 
if 
some amount of fine tuning is allowed.
\cite{Cacciapaglia:2004rb}
It is an interesting 
problem 
to determine whether 
boundary conditions that
reduce 
the rank of the group
can be 
forbidden as a limit of the discretized gauge theories 
as in the deconstruction approach \cite{Arkani-Hamed:2001ca}. 

It is also an interesting 
problem to build a realistic 
model using the 5D 
gauge theories with 
boundaries. 
Left-right symmetric models have been 
extensively studied in 
model building \cite{Csaki:2005vy}. 
It 
would be interesting to obtain 
a reduction of the rank of 
the gauge group with such a left-right symmetric model. 
Although they 
employ linear combinations of generators of 
the factor group, 
it should  be examined carefully 
whether the 
rank reduction is 
indeed
compatible with 
the restricted class of 5D 
gauge transformations.

%
%

\vspace*{10mm}
\section*{Acknowledgements}
We wish to thank Nobuhito Maru and Makoto Sakamoto for 
useful discussions, and Masaharu Tanabashi for 
discussions as 
well as for informing us of his unpublished results and 
literature 
on the deconstruction approach. 
This work is supported in part by Grants-in-Aid for Scientific 
Research from the Ministry of Education, Culture, Sports, 
Science and Technology 
of Japan [Nos.17540237 (N.S.) 
and 16028203 for the 
Priority 
Area ``
Origin of 
Mass'' 
(N.S.~and N.U.)]. 


\begin{appendix}
\section{Formulas for Mode Functions and Overlap Integrals\label{Ap:KK}}
In this appendix, we summarize properties of the mode 
functions $f_n^{D(a)}(y)$ and $g_m^{D(b)}(y)$ and overlap 
functions 
constructed from them.

The mode functions $f_n^{D(a)}$ and $g_n^{D(a)}$ are 
defined by the eigenvalue equations with the mass 
eigenvalue $M_n$ as 
\begin{eqnarray}
 -{1\over \sqrt{g_{55}}}\partial_5 {e^{-4W}\over \sqrt{g_{55}}}
   \partial_5 f_n^{D(a)}(y)
&=&M_n^2 f_n^{D(a)}(y) ,
\\ 
 -\partial_5 {1\over \sqrt{g_{55}}}\partial_5
    {e^{-4W}\over \sqrt{g_{55}}} g_n^{D(a)}(y)&=&M_n^2 g_n^{D(a)}(y) ,
\end{eqnarray}
which can be rewritten 
as the following 
coupled first-order differential equations 
\begin{equation}
 \partial_5 f_n^{D(a)}=M_ng_n^{D(a)},
 ~~~~~
-{1\over \sqrt{g_{55}}}\partial_5 {e^{-4W}\over
  \sqrt{g_{55}}}g_n^{D(a)}=M_n f_n^{D(a)} .
\label{eq:coupled_eq}
\end{equation}
We choose them to be orthonormal, 
\begin{eqnarray}
 \int_0^{\pi R} dy \sqrt{g_{55}} f_n^{D(a)}f_m^{D(a)}=\delta_{nm} ,
 &&
 \int_0^{\pi R} dy \sqrt{g_{55}}^{-1}e^{-4W}
    g_n^{D(a)}g_m^{D(a)} 
  =\delta_{nm} ,
\end{eqnarray}
and assume them to be complete 
\begin{eqnarray}
\sum_n f_n^{D(a)}(y) f_n^{D(a)}(y')&=&{1\over \sqrt{g_{55}}}\delta(y-y') ,
\\
\sum_n g_n^{D(a)}(y) g_n^{D(a)}(y')&=&e^{4W}\sqrt{g_{55}}\delta(y-y')   .
\end{eqnarray}

Using the KK decompositions in 
Eqs.(\ref{eq:kk_vector}) and (\ref{eq:kk_scalar}) and 
integrating over $y$, 
we obtain the action for the field strengths without the 
gauge fixing term in Eq.(\ref{eq:lagr}) as 
\begin{eqnarray}
 {\cal S}&\!\!\!=&\!\!\!\int d^4x \bigg[
  -{1\over 4}(\partial_{\mu}A_{\nu n}^a-\partial_{\nu}A_{\mu
  n}^a)
     (\partial^{\mu}A_n^{\nu a}-\partial^{\nu}A_n^{\mu a})
    -{1\over 2} M_n^2 A_{\mu n}^a A_n^{\mu a}
 +M_n(\partial_{\mu} A_{5n}^a)A_n^{\mu a}
\nonumber
\\ 
   &&\qquad
    -g_5 f^{abc}A_{\mu n}^b A_{\nu m}^c \partial^{\mu}A_l^{\nu a}
    F_{nml}^{D(bca)}
   -{1\over 4}g_5^2 f^{abc}f^{ade}A_{\mu m}^b A_{\nu n}^c
    A_l^{\mu d}A_k^{\nu e}F_{mnlk}^{D(bcde)} 
\nonumber
\\
  &&\qquad
  -{1\over 2}\partial_{\mu}A_{5n}^a \partial^{\mu} A_{5n}^a
   -g_5 f^{abc}A_{\mu n}^b A_{5m}^c \partial^{\mu}A_{5l}^a
     T_{nml}^{D(bca)}
\nonumber
\\ 
 &&\qquad
  +M_l g_5 f^{abc}A_{\mu n}^b A_{5m}^c A_l^{\mu a}
   T_{nml}^{D(bca)}
  -{1\over 2}g_5^2 f^{abc}f^{ade}A_{\mu m}^b A_{5 n}^c
       A_l^{\mu d}A_{5 k}^e T_{mnlk}^{D(bcde)}
 \bigg] \!,
  \label{KKaction}
\end{eqnarray}
where the overlap integrals are given by 
\begin{eqnarray}
 F_{nml}^{D(bca)}&=&\int_0^{\pi R} dy
  \sqrt{g_{55}}f_n^{D(b)}f_m^{D(c)}f_l^{D(a)} ,
\nonumber
  \\
T_{nml}^{D(bca)}&=&\int_0^{\pi R} dy {e^{-4W}\over \sqrt{g_{55}}}
     f_n^{D(b)} g_m^{D(c)} g_l^{D(a)} ,
\nonumber
\\
   F_{mnlk}^{D(bcde)}
  &=&\int_0^{\pi R} dy \sqrt{g_{55}}f_m^{D(b)} f_n^{D(c)} f_l^{D(d)}
     f_k^{D(e)} ,
\nonumber
\\
 T_{mnlk}^{D(bcde)}
     &=&\int_0^{\pi R} dy {e^{-4W}\over\sqrt{g_{55}}}
  f_m^{D(b)}g_n^{D(c)} f_l^{D(d)} g_k^{D(e)} .
\label{overlap}
\end{eqnarray}

By repeatedly using the defining equations 
(\ref{eq:coupled_eq}), we obtain 
the 
useful identity 
\begin{eqnarray}
&& M_m(M_n T_{mmn}^{D(bda)}-M_m T_{nmm}^{D(abd)})
  =(M_n^2-M_m^2)F_{nmm}^{D(abd)}
\nonumber
\\    
& &+\left[{e^{-4W}\over \sqrt{g_{55}}}(f_m^{D(b)} f_m^{D(d)}
       f_n^{D(a)}{}'
         -f_n^{D(a)} f_m^{D(b)} f_m^{D(d)}{}')\right]_0^{\pi R} . 
 \label{mtb}
\end{eqnarray}

\section{Calculation of Scattering Amplitude \label{Ap:amp}}

\begin{figure}[h]
\begin{center}
\includegraphics{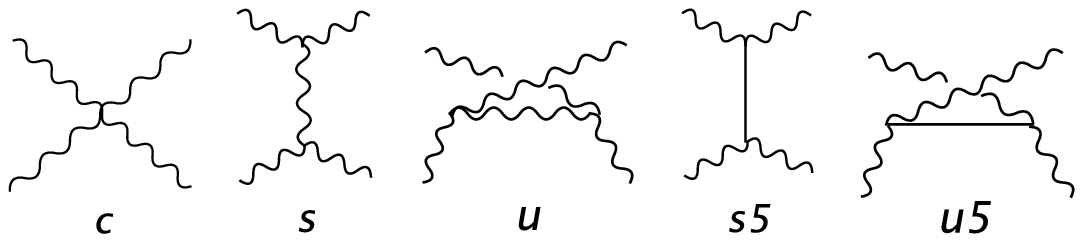}
\end{center}
\caption{The tree level diagrams: The  $A_\mu$ and $A_5$ components 
of the gauge bosons are 
represented by wavy lines and straight lines, 
respectively.\label{tree}} 
\end{figure}

In this appendix, we give some details 
concerning the calculation of  
the scattering amplitude $A_n^a A_m^b\to A_l^c A_m^d$ at 
tree 
level with the high energy approximation, 
in which the
total energy $E$ is sufficiently large 
compared to the mass 
of any external 
or intermediate state. 
The tree level diagrams 
are 
presented\footnote{ 
Because we are interested in the case 
in which the gauge 
boson $A_l^c$ can have a zero mode  ($l=0$),
we 
list 
only the diagrams appropriate for that case. 
} 
in Fig.~\ref{tree}.

\begin{figure}[h]
\begin{center}
\includegraphics[height=3.5cm]{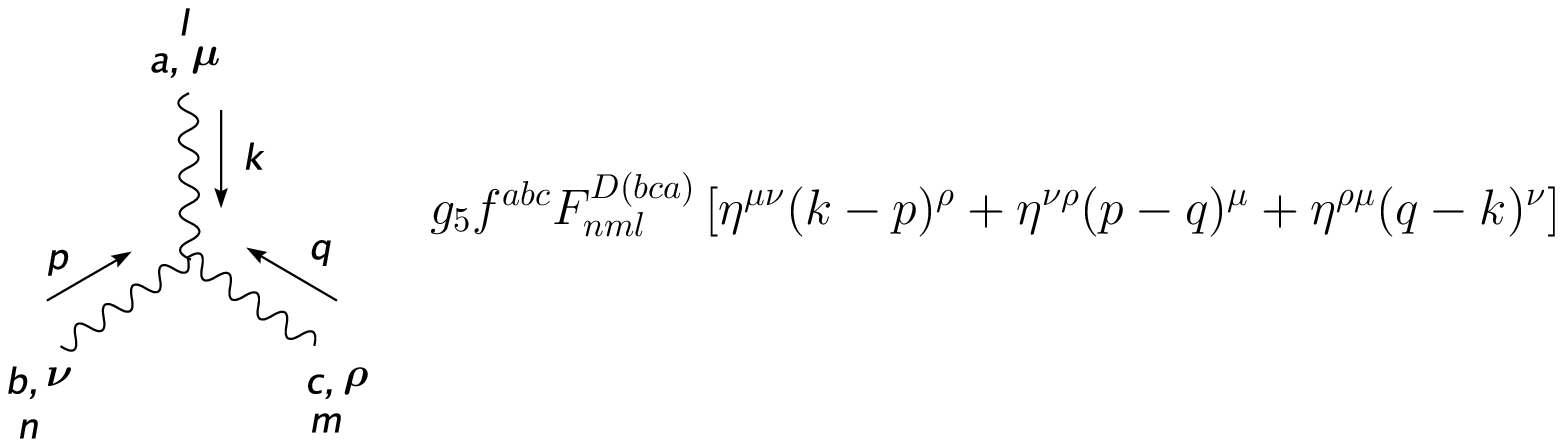} \\
\includegraphics[height=3cm]{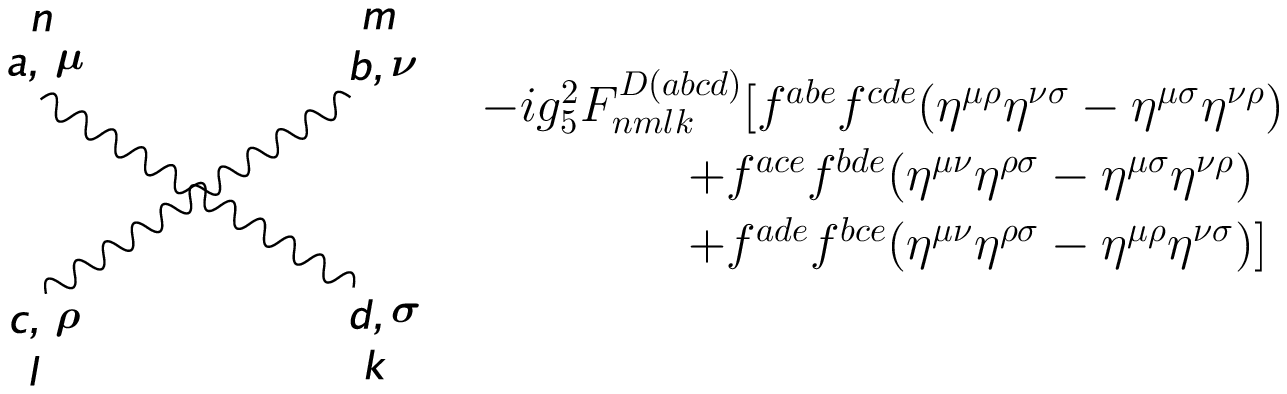} \\
\includegraphics[height=3cm]{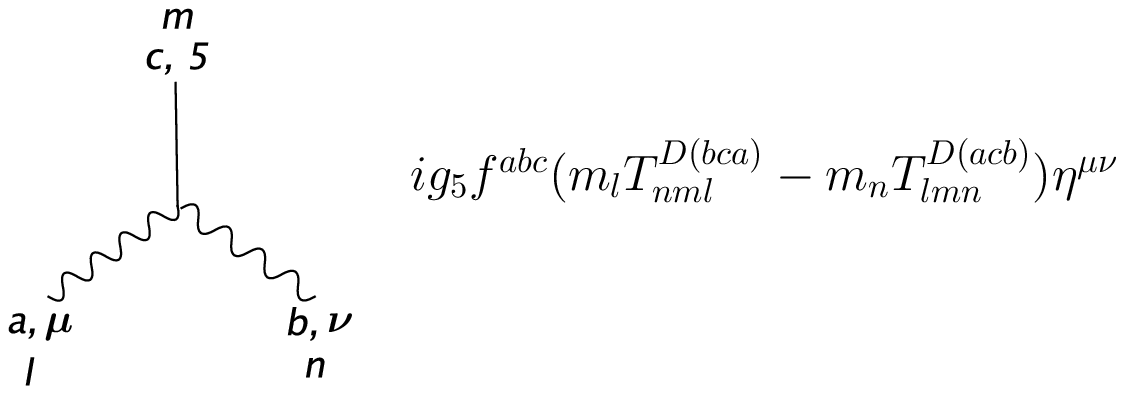}
\end{center}
\caption{Vertices. \label{vertex}} 
\end{figure}

In the $R_{\xi}$ gauge, the propagators for $A_{\mu}$ and $A_5$ 
have a $\xi$ dependent mass: 
\begin{eqnarray}
&&
\left\langle A_{\mu m}^a A_{\nu k}^b\right\rangle
    ={-i\delta^{ab}\delta_{mk}\over p^2+M_k^2}
  \left(\eta_{\mu\nu}-(1-\xi){p_{\mu}p_{\nu}\over p^2 + \xi
     M_k^2}\right) , 
\\
  && \left\langle A_{5 m}^a A_{5 k}^b\right\rangle
    ={-i\delta^{ab}\delta_{mk}\over p^2+\xi M_k^2} .
  \label{a5prop}
\end{eqnarray}
The 
vertices can be read off 
of the KK decomposed 
action (\ref{KKaction}), 
as given in Fig.~\ref{vertex}.
We use 
an expansion in powers of $M_k/E$. 
As seen from the propagators, this expansion requires 
the condition
$\xi\ll E^2/M_k^2$, which is possible except in the 
unitary gauge, where we have $\xi \to \infty$.

Assuming $\xi\ll E^2/M_k^2$, 
the contribution 
of each diagram in Fig.\ref{tree}
to the invariant matrix element 
is listed in Table~\ref{Tab:xi1}. 
\begin{table}[b]
\caption{Contribution from each diagram to 
the invariant matrix element 
in the case $\xi\ll E^2/M_k^2$.
Each contribution should be multiplied by the factor 
$-ig_5^2 f^{abe}f^{cde} E^3/(8M_n M_m^2)\epsilon_{\mu}^*(p_3)$.
Here, 
the Jacobi identity 
$f^{ade}f^{cbe}=f^{abe}f^{cde}$ 
has been used.\label{Tab:xi1}}
\begin{eqnarray}
c &\Bigg|&
 F_{lmnm}^{D(cdab)}\left(1-{M_l^2\over E^2}\right)
 \Bigg(1-{M_m^2\over E^2}+\left(3-{4M_n^2-M_m^2\over
			     E^2}\right)\cos\theta,
   -2\left(1-{M_n^2\over E^2}\right)\sin\theta,
\nonumber\\
  &&\qquad\qquad\qquad\qquad
  0,-3+{2M_n^2+7M_m^2\over E^2}
    -\left(1-{2M_n^2+M_m^2\over E^2}\right)\cos\theta\Bigg)^{\mu} .
\nonumber\\
s &\Bigg|&
  -2\sum_k F_{lmk}^{D(cde)}F_{nmk}^{D(abe)} \left(1-{M_l^2\over E^2}\right)
  \Bigg({M_n^2-M_m^2\over E^2}+4{M_m^2\over E^2}\cos\theta,
     {M_n^2-M_m^2\over E^2}\sin\theta ,
\nonumber\\
  &&\qquad\qquad\qquad\qquad
   0,-2\left(1-{M_m^2-M_k^2\over E^2}\right)
     +{M_n^2-M_m^2\over E^2}\cos\theta \Bigg)^{\mu} .
\nonumber\\
u &\Bigg|&
  -\sum_k F_{mnk}^{D(dae)}F_{mlk}^{D(bce)}\left(1-{M_l^2\over
					   E^2}\right)
\nonumber
\\
  && \times
   \Bigg(2+{M_n^2-2M_m^2+3M_l^2-4M_k^2\over E^2}
     +\left(2-{5M_n^2+8M_m^2+M_l^2\over E^2}\right)\cos\theta,
\nonumber\\
  &&
     -\left(1-{M_n^2-5M_m^2-M_l^2+2M_k^2\over E^2}\right)\sin\theta
      -\left(1+{M_n^2+M_m^2-M_l^2\over
	E^2}\right)\sin\theta\cos\theta,0,
\nonumber\\
  &&
    1+{4M_n^2+5M_m^2+2M_l^2-2M_k^2\over E^2}
   +\left({M_n^2-4M_m^2-M_l^2+2M_k^2\over E^2}\right)\cos\theta
\nonumber\\
  &&
    -\left(1+{M_n^2+M_m^2-M_l^2\over E^2}\right)\cos^2\theta\Bigg)^{\mu} .
\nonumber\\
s5 &\Bigg|&
 -{2\over E^2}\sum_k \left(M_l T_{mkl}^{D(dec)}-M_m T_{lkm}^{D(ced)}\right)
    \left(M_n T_{mkn}^{D(bea)}-M_m T_{nkm}^{D(aeb)}\right)
      \left(1,-\sin\theta, \right.
\nonumber\\
  &&\qquad\qquad\qquad\qquad \left.
 0,-\cos\theta\right)^{\mu} .
\nonumber\\
u5 &\Bigg|&
 -{2\over E^2}\sum_k \left(M_n T_{mkn}^{D(dea)}-M_m T_{nkm}^{D(aed)}\right)
    \left(M_l T_{mkl}^{D(bec)}-M_m T_{lkm}^{D(ceb)}\right)
      \left(-1,0,0,1\right)^{\mu} .
\nonumber
\end{eqnarray}
\end{table}
Summing all the contributions in Table~\ref{Tab:xi1},
we obtain the total invariant matrix element as
\begin{eqnarray}
 && ig_5^2 f^{abe}f^{cde} {E^3\over 8M_n M_m^2}\epsilon_{\mu}^*(p_3)
 \sum_k \Bigg\{ F_{lmk}^{D(cde)}F_{nmk}^{D(abe)}
 {\cal A}^{\mu}
\nonumber\\
&&
 +{2\over E^2} \left(M_l T_{mkl}^{D(dec)}-M_m T_{lkm}^{D(ced)}\right)
    \left(M_n T_{mkn}^{D(bea)}-M_m T_{nkm}^{D(aeb)}\right)
      \left(0,-\sin\theta,0,1-\cos\theta\right)^{\mu}
   \Bigg\}
\nonumber\\   \label{ime}
\end{eqnarray}
with 
\begin{eqnarray}
 {\cal A}^{\mu}
 &\!=\!& \Bigg(1+{3M_n^2-3M_m^2+2M_l^2-4M_k^2\over E^2}
    -\left(1+{M_n^2+M_m^2\over E^2}\right)\cos\theta,
\nonumber\\
&&\quad
    \left(1+{M_n^2-7M_m^2-2M_l^2+2M_k^2\over E^2}\right)\sin\theta
    -\left(1+{M_n^2+M_m^2-2M_l^2\over E^2}\right)\sin\theta\cos\theta,
    0,
\nonumber\\
&&\quad
     {2M_n^2+4M_m^2+2M_l^2-6M_k^2\over E^2}
     +\left(1+{M_n^2-7M_m^2-2M_l^2+2M_k^2\over E^2}\right)\cos\theta
\nonumber\\
&&\quad
     -\left(1+{M_n^2+M_m^2-2M_l^2\over
       E^2}\right)\cos^2\theta\Bigg)^{\mu} ,
 \label{cala}
\end{eqnarray}
where $F_{lmnm}^{D(cdab)}=\sum_k F_{lmk}^{D(cde)}F_{nmk}^{D(abe)}$.
We have not yet specified the polarization $\epsilon_\mu(p_3)$ 
of the  external $A_l^c$ boson 
so that we are free to 
choose 
a transverse or longitudinal polarization as well as 
a massless or massive $A_l^c$ 
in the following.

Let us first consider the case 
in which the $A_l^c$ boson 
is a massive KK mode.
Then, the on-shell polarization can be transverse, as in 
Eq.~(\ref{polart}), or longitudinal, as in Eq.~(\ref{polarl}). 
If we choose 
the transverse polarization 
$\epsilon^*(p_3)=(0,\cos\theta,0,-\sin\theta)$, 
we find that the 
invariant matrix element (\ref{ime}) reduces to 
Eq.~(\ref{tk}), whereas the other transverse polarization, 
$\epsilon^*(p_3)=(0,0,1,0)$, makes the invariant matrix 
element vanish. 
If we choose the longitudinal 
polarization in Eq.~(\ref{polarl}), 
we find that the 
invariant matrix element (\ref{ime}) reduces to 
Eq.~(\ref{lk}).

Let us next consider the case 
in which the external 
$A_l^c$ boson is massless. 
Then, only the state with $M_k=M_m$ is allowed in the 
intermediate state, 
since $F_{lmk}^{D(cde)}$ and
$T_{lkm}^{D(ced)}$ reduce to $f_l^{D(c)}\delta_{mk}$. 
Before taking the high energy limit, 
we obtain the result for arbitrary $\xi$ by 
substituting the polarization of $A_l^c$ with its momentum, 
\begin{eqnarray}
 &&ig_5^2 f^{abe}f^{cde} \times
\\
 &&\bigg[
   f_l^{D(c)}\{(M_n^2-M_m^2)F_{nmm}^{D(abd)}
     -M_m(M_n T_{mmn}^{D(bda)}-M_m T_{nmm}^{(abd)})\}
     {p_3^\sigma\eta^{\mu\nu}\over (p_3+p_4)^2+\xi M_m^2}
\nonumber\\
 && + f_l^{D(c)}\{(M_n^2-M_m^2)F_{nmm}^{D(abd)}
     -M_m(M_n T_{mmn}^{D(dba)}-M_m T_{nmm}^{(abd)})\}
     {p_3^\nu\eta^{\mu\sigma}\over (p_4-p_1)^2+\xi M_m^2}
 \bigg]. 
\nonumber
\end{eqnarray}
Equation (\ref{mtb}) implies that this amplitude is proportional 
to a function of the values of the mode functions at the 
boundaries.  
Moreover, it has $\xi$ in 
its denominator. 
Therefore, the unitary gauge, for which $\xi \to \infty$, clearly 
misses this possible violation of the Ward-Takahashi identity. 
Returning to the high energy approximation, we find 
by using Eq.~(\ref{mtb}) that 
the invariant matrix element in Eq.~(\ref{ime}) 
reduces to Eq.~(\ref{eq:ward_takahashi_id}).

\end{appendix}

\vspace*{10mm}


\end{document}